\documentclass[twocolumn,prc,showpacs,preprintnumbers,amsmath,amssymb]{revtex4}
\usepackage{graphicx}% Include figure filesph
\usepackage{dcolumn}% Align table columns on decimal point
\usepackage{bm}% bold math
\usepackage{times}

\def\dfrac#1#2{{\displaystyle\frac{#1}{#2}}}
\def\bs{\boldsymbol}
\def\mr{\mathrm}

\def\non{\nonumber}

\def\beq{\begin{equation}}
\def\eeq{\end{equation}}
\def\bea{\begin{eqnarray}}
\def\eea{\end{eqnarray}}
\begin{document}

\preprint{}

\title{
Off-the-energy-shell $pp$ scattering in the exclusive proton knockout
$^{12}$C($p,2p$) reaction at 392 MeV
}

\author{T.~T. Sumi}
\affiliation{Department of Physics, Kyushu University,
Fukuoka 812-8581, Japan}
\author{K. Ogata}
\email[Electronic address: ]{kazu2scp@kyudai.jp}
\affiliation{Department of Physics, Kyushu University,
Fukuoka 812-8581, Japan}

\date{\today}

\begin{abstract}
The triple differential cross section (TDX) for the $^{12}$C($p,2p$) proton
knockout reaction from the 1p$3/2$ single-particle state measured
at 392 MeV is investigated by nonrelativistic distorted wave impulse
approximation (DWIA) with accurate
treatment of the kinematics of the colliding two protons
and effects of in-medium modification to the matrix elements of
the proton-proton ($pp$) effective interaction.
Some simplifying approximations made in previous studies with DWIA
are examined.
The off-the-energy-shell matrix elements of the
$pp$ effective interaction
are shown to play an essential role in describing the
asymmetric two peaks of the measured TDX corresponding to
the kinematics in which the momentum transfer is fixed.
\end{abstract}


\maketitle

Nucleon knockout reaction is  a powerful tool to extract information
on single-particle states of nuclei.
Proton induced knockout reaction
such as ($p,2p$) reaction at intermediate energies \cite{Redish,p2p1,p2p2}
has been studied intensively
because of its significantly large cross section compared
to the electron induced ($e,e'p$) one.
A compensation for the large cross section is that the interaction concerned
is not understood sufficiently.
In fact, it is known that spin observables of the ($p,2p$) reaction,
the analyzing power $A_y$ in particular, cannot be explained by
standard
distorted wave impulse approximation (DWIA) calculations
\cite{nrDWIA,AyPuzzle2}.
Recently, this discrepancy has been attributed
to effects of in-medium modification to the proton-proton ($pp$) effective
interaction; the reduction of masses of mesons in nuclear medium \cite{BR}
was shown to improve the agreement between experimental
data of $A_y$ and results of theoretical calculation within the
framework of relativistic impulse
approximation (RIA) \cite{Noro1,Hillhouse1,Noro2,Hillhouse2}.
The reduction of meson masses,
however, brought about a serious problem on other spin observables
such as the spin transfer coefficients $D_{ij}$ \cite{Noro1}.

On the other hand, effects of in-medium modification to the
matrix elements of the effective $pp$ interaction,
henceforth called in-medium effects,
on observables of the ($p,2p$) reaction within the nonrelativistic
framework, have not been investigated well. In particular,
the off-the-energy-shell (off-shell) property of the $pp$ scattering
in nuclear medium was not treated properly in the foregoing studies
with nonrelativistic DWIA \cite{nrDWIA,AyPuzzle2}.
Moreover, effects of refraction, i.e., change
in the kinematics of the incoming and outgoing protons caused by
distortion of the target nucleus, were also neglected.
Since the refractive effect was shown to
be essential to reproduce $A_y$ data of inclusive ($p,p'x$) reactions
\cite{Ogata},
it is expected to be significant in the calculation of
spin observables of also ($p,2p$) reactions. Careful analysis of the
($p,2p$) reactions with nonrelativistic DWIA taking account of
the above-mentioned effects is necessary to draw a conclusion on
the comparison between nonrelativistic calculation and experimental
data of ($p,2p$) reactions.

As the first step to such theoretical analysis of ($p,2p$) reactions,
in this Rapid Communication we analyze the triple differential
cross sections (TDX's) of the $^{12}$C($p,2p$) reaction at 392 MeV
corresponding to
the knockout of proton in the 1p$3/2$ single-particle state in $^{12}$C.
Our main purpose of the present study is to show that the off-shell $pp$
scattering is essential to reproduce the experimental data of the TDX
for specific kinematics in which the momentum transfer is fixed.

We start with the usual DWIA $T$ matrix of the following form:
\bea
T_{m'_{1}m'_{2}m_1M_j}
&=&
\iint d{\bs r}_{1}d{\bs r}_{2}d\xi_1 d\xi_2
\bar{\chi}_{\mr{f}_{1};{\bs k}'_{1},m'_{1}}^{(-)*}
({\bs r}_{1},\xi_1)
\non \\
&&
\times
\bar{\chi}_{\mr{f}_{2};{\bs k}'_{2},m'_{2}}^{(-)*}
({\bs r}_{2},\xi_2)
\tau (\bs{s};\rho({\bs R}),K_{NN})
\non \\
&&
\times
\bar{\chi}_{\mr{i};{\bs k}_{1},m_{1}}^{(+)}
({\bs r},\xi_1)
\bar{\phi}_{1p \frac{3}{2},M_j}({\bs r}_{2},\xi_2),
\label{Tmat}
\eea
where
$\bar{\chi}_{\mr{f}_1}$ ($\bar{\chi}_{\mr{f}_2}$) is the distorted wave for
the outgoing proton 1 (proton 2) relative to the residual nucleus
$^{11}$B;
$m_1$ ($m_2$) and ${\bs k}'_1$ (${\bs k}'_2$)
are, respectively, the $z$-component of the spin
and the asymptotic momentum in $\hbar$ unit
of the observed proton 1 (proton 2).
The coordinate of the proton $i$ ($i=1$, 2) relative to
the residual nucleus $^{11}$B is denoted by ${\bs r}_i$, and
$\xi_i$ is the corresponding internal coordinate.
The distorted wave between the incident proton and $^{12}$C target
is denoted by $\bar{\chi}_{\mr{i}}$ with
the asymptotic momentum ${\bs k}_1$ and
the $z$-component $m_1$ of the spin of the proton in the asymptotic
region.
The relative coordinate $\bs{r}$ for $\bar{\chi}_{\mr{i}}$ is given by
${\bs r}_{1}-{\bs r}_{2}/A$
with the mass number $A$ of the target nucleus.
Furthermore, $\bar{\phi}_{1p\frac{3}{2},M_j}$ is
the single-particle wave function of the target
proton in the 1p3/2 state and $M_j$ is the $z$-component of the total
angular momentum $j$ that is 3/2 in the present case.
$\tau$ is the $pp$ effective interaction in coordinate representation
with ${\bs s}={\bs r}_1-{\bs r}_2$
and $K_{NN}$ the scattering energy of the two proton system explained below
in more detail.
Note that $\tau$ contains
the antisymmetrization operator for the colliding two protons.
We use in the present study the half-off-shell $g$
matrix developed by the Melbourne
group \cite{Amos} based on Bonn-B nucleon-nucleon ($NN$) potential \cite{BonnB}, the Melbourne $g$ matrix, as for $\tau$, which depends on
also the nuclear density $\rho$; we adopt $\rho$ at ${\bs R}=({\bs r}_1
+{\bs r}_2)/2$
on the basis of local density approximation.
The $T$-matrix elements of Eq.~(\ref{Tmat}) are evaluated in the center of mass
(c.m.) frame of the total reaction system with
the relativistic kinematics.
The TDX is given by
\beq
\dfrac{d^3\sigma}
{dE_1^\mr{L}
d\Omega_1^\mr{L}
d\Omega_2^\mr{L}}
=
\dfrac{JF_\mr{kin}S}{2(2j+1)}
\sum_{m'_1m'_2m_1M_j}
\!\!
\left|
T_{m'_{1}m'_{2}m_1M_j}
\right|^2,
\label{TDX}
\eeq
where $\Omega_1^\mr{L}$ ($\Omega_2^\mr{L}$) is the solid angle for the emitted
proton 1 (proton 2) and $E_1^\mr{L}$ is the emission energy of the
proton 1;
the superscript L denotes the laboratory frame.
$J$ is the Jacobian
for the transformation from the c.m. frame to the
laboratory one,
$F_\mr{kin}$ is the kinetic factor
and $S$ is the spectroscopic factor
that agrees with $2j+1=4$ if the single-particle description of
the target nucleus is fully accurate.

We in the present study neglect the spin dependent part of the
distorting potentials for the $\bar{\chi}$, which is known to give no
significant effects on unpolarized cross sections.
We then have
\bea
\bar{\chi}_{\mr{f}_{1};{\bs k}'_{1},m'_{1}}^{(-)}({\bs r}_{1},\xi_1)
&\to&
\chi_{\mr{f}_{1};{\bs k}'_{1}}^{(-)}({\bs r}_{1})
\eta_{m'_{1}}(\xi_1),
\non \\
\bar{\chi}_{\mr{f}_{2};{\bs k}'_{2},m'_{2}}^{(-)}({\bs r}_{2},\xi_2)
&\to&
\chi_{\mr{f}_{2};{\bs k}'_{2}}^{(-)}({\bs r}_{2})
\eta_{m'_{2}}(\xi_2),
\non \\
\bar{\chi}_{\mr{i};{\bs k}_{1},m_{1}}^{(+)}
({\bs r},\xi_1)
&\to&
\chi_{\mr{i};{\bs k}_{1}}^{(+)}({\bs r})
\eta_{m_{1}}(\xi_1),
\non
\eea
where the $\chi$ are the spatial parts of the $\bar{\chi}$ and
$\eta$ is the spin 1/2 wave function.
The spin dependence of $\bar{\phi}_{1p\frac{3}{2},M_j}$
is taken into account with
\beq
\bar{\phi}_{1p\frac{3}{2},M_j}({\bs r}_{2},\xi_2)
=
\sum_{m,m_2}(1m{\textstyle\frac{1}{2}}m_2|{\textstyle\frac{3}{2}} M_j)
\phi_{1p\frac{3}{2},m}(\bs{r}_2)
\eta_{m_2}(\xi_2),
\eeq
where $\phi_{1p\frac{3}{2},m}$ is the spatial
part of $\bar{\phi}_{1p\frac{3}{2},M_j}$
with $m$ the $z$-component of the orbital angular momentum.

The relative coordinates
concerned with the $\chi$ are expressed with $\bs{R}$ and $\bs{s}$
as follows:
\bea
\bs{r}_1 &=& {\bs R} + {\bs s}/2,
\non \\
{\bs r}_2 &=& {\bs R} - {\bs s}/2,
\non \\
{\bs r} &=&
 \dfrac{A-1}{A} {\bs R} + \dfrac{A+1}{2A} {\bs s}
\equiv
 \eta {\bs R} + \zeta {\bs s}.
\non
\eea
We now make the local semi-classical approximation (LSCA) \cite{Luo}
to the $\chi$, i.e.,
\beq
\chi({\bs u}+{\bs v})
\approx
\chi({\bs u})
\exp[i{\bs k}({\bs u})\cdot{\bs v}].
\eeq
The magnitude of the local momentum ${\bs k}({\bs u})$ is evaluated by
the real part of the complex local momentum that satisfies
the local energy conservation law at ${\bs u}$. The direction
of ${\bs k}({\bs u})$ is taken
to be parallel with the flux of $\chi$ at ${\bs u}$. The LSCA
is shown to be valid for $|{\bs v}|$ smaller than about 1.5 fm
\cite{Watanabe}.
The LSCA is applied to each distorted wave in the $T$ matrix of
Eq.~(\ref{Tmat}) as
\bea
\chi_{\mr{f}_{1};{\bs k}'_{1}}^{(-)}
({\bs R} + {\bs s}/2)
&\approx&
\chi_{\mr{f}_{1};{\bs k}'_{1}}^{(-)}
({\bs R})
\exp[i{\bs k}'_1({\bs R})\cdot{\bs s}/2],
\label{LSCA1}
\\
\chi_{\mr{f}_{2};{\bs k}'_{2}}^{(-)}
({\bs R} - {\bs s}/2)
&\approx&
\chi_{\mr{f}_{2};{\bs k}'_{2}}^{(-)}
({\bs R})
\exp[-i{\bs k}'_2({\bs R})\cdot{\bs s}/2],
\label{LSCA2}
\\
\chi_{\mr{i};{\bs k}_{1}}^{(+)}
(\eta {\bs R} + \zeta {\bs s})
&\approx&
\chi_{\mr{i};{\bs k}_{1}}^{(+)}
(\eta {\bs R})
\exp[i{\bs k}_1(\eta{\bs R})\cdot \zeta{\bs s}].
\label{LSCA3}
\eea
Note that this approximation to the $\chi$ is
valid for $s$ less than about 3 fm.
Since the integrand of Eq.~(\ref{Tmat}) is appreciable for small
$s$ less than the range of $\tau$, which is about 1.5 fm,
one can evaluate with high accuracy the $T$ matrix making use of
Eqs.~(\ref{LSCA1})--(\ref{LSCA3}). One thus obtains
\bea
T_{m'_{1}m'_{2}m_1M_j}
&\approx&
\int d{\bs R}
\chi_{\mr{f}_{1};{\bs k}'_{1}}^{(-)*}({\bs R})
\chi_{\mr{f}_{2};{\bs k}'_{2}}^{(-)*}({\bs R})
\chi_{\mr{i};{\bs k}_{1}}^{(+)}(\eta {\bs R})
\non \\
&&
\times
\sum_{m,m_2}
(1m{\textstyle\frac{1}{2}}m_2|{\textstyle\frac{3}{2}} M_j)
\int d{\bs s}d\xi_1 d\xi_2
\non \\
&&
\times
\eta^*_{m'_1}(\xi_1)
\eta^*_{m'_2}(\xi_2)
\non \\
&&
\times
\exp\left[-i\dfrac{{\bs k}'_1({\bs R}) - {\bs k}'_2({\bs R})}{2}
\cdot{\bs s}\right]
\non \\
&&
\times
\tau (\bs{s};\rho({\bs R}),K_{NN})
\non \\
& &
\times
\exp\left[i\zeta{\bs k}_1(\eta{\bs R})\cdot {\bs s}\right]
\eta_{m_1}(\xi_1)
\eta_{m_2}(\xi_2)
\non \\
&&
\times
\phi_{1p\frac{3}{2},m}({\bs R} - {\bs s}/2).
\label{Tmat2}
\eea

The spatial part of the
proton single-particle wave function
$\phi_{1p\frac{3}{2},m}$ can be written with
its Fourier transform $\tilde{f}$ as
\bea
\phi_{1p\frac{3}{2},m}({\bs R} - {\bs s}/2)
&=&
\dfrac{1}{(2\pi)^3}
\int
d\bs{k}_2^\mr{A}
\tilde{f}_{1p\frac{3}{2},m}(\bs{k}_2^\mr{A})
\non \\
& &
\times
\exp[i\bs{k}_2^\mr{A}\cdot({\bs R} - {\bs s}/2)],
\label{FT}
\eea
where $\bs{k}_2^\mr{A}$ can be interpreted as the relative
momentum between the target proton and the residual nucleus in the
c.m. frame of the target nucleus $^{12}$C.
Inserting Eq.~(\ref{FT}) into Eq.~(\ref{Tmat2}) one obtains
\bea
T_{m'_{1}m'_{2}m_1M_j}
&\approx&
\dfrac{1}{(2\pi)^3}
\int d{\bs R}
\chi_{\mr{f}_{1};{\bs k}'_{1}}^{(-)*}({\bs R})
\chi_{\mr{f}_{2};{\bs k}'_{2}}^{(-)*}({\bs R})
\chi_{\mr{i};{\bs k}_{1}}^{(+)}(\eta {\bs R})
\non \\
&&
\times
\sum_{m,m_2}
(1m{\textstyle\frac{1}{2}}m_2|{\textstyle\frac{3}{2}} M_j)
\int
d\bs{k}_2^\mr{A}
\tilde{f}_{1p\frac{3}{2},m}(\bs{k}_2^\mr{A})
\non \\
&&
\times
e^{i\bs{k}_2^\mr{A}\cdot{\bs R}}
M_{m'_1m'_2m_1m_2}(\bs{\kappa}',\bs{\kappa}),
\label{Tmat3}
\eea
where the matrix elements $M$ of $\tau$ are defined by
\bea
M_{m'_1m'_2m_1m_2}(\bs{\kappa}',\bs{\kappa})
&\equiv&
\int d{\bs s}d\xi_1 d\xi_2
\eta^*_{m'_1}(\xi_1)
\eta^*_{m'_2}(\xi_2)
\non \\
&&
\times
e^{-i{\bs \kappa}'({\bs R})\cdot{\bs s}}
\tau (\bs{s};\rho({\bs R}),K_{NN})
\non \\
& &
\times
e^{i{\bs \kappa}(\eta{\bs R})\cdot {\bs s}}
\eta_{m_1}(\xi_1)
\eta_{m_2}(\xi_2)
\label{ME}
\eea
with
\bea
{\bs \kappa}'&=&
\dfrac{{\bs k}'_1({\bs R}) - {\bs k}'_2({\bs R})}{2},
\label{kpf} \\
{\bs \kappa} &=&
\dfrac{2\zeta{\bs k}_1(\eta{\bs R})-\bs{k}_2^\mr{A}}{2}.
\non
\eea
One immediately sees ${\bs \kappa}'$ is the relative momentum between
the two protons in the final state of the transition by $\tau$
at ${\bs R}$ in the c.m. frame of the $p+^{12}$C system.
For the initial state, since the motion
of the target proton can be described  well with nonrelativistic
kinematics, one may approximate
the relative momentum ${\bs \kappa}$ as
\bea
{\bs \kappa}
&\approx&
\dfrac{
{\bs k}_1(\eta{\bs R})
-
{\bs k}_2
+
\{
{\bs k}_1(\eta{\bs R})
-
{\bs k}_1
\}/A
}
{2}
\non \\
&\approx&
\dfrac{
{\bs k}_1(\eta{\bs R})
-
{\bs k}_2
}
{2},
\label{kpi}
\eea
where
${\bs k}_2 \equiv \bs{k}_2^\mr{A}-{\bs k}_1/A$ is to be interpreted
as the momentum of the target proton in the initial state
in the $p+^{12}$C c.m. frame.
%%%Note that ${\bs k}_1$
%%%is the asymptotic (not local) momentum of the incoming proton.

Thus, for given $\bs{R}$ and $\bs{k}_2^\mr{A}$ in Eq.~(\ref{Tmat3}),
the kinematics of the colliding two protons corresponding to
the initial and the final states of the transition by $\tau$
is uniquely determined.
Specification of the momentum of each proton in the initial state
is necessary since the scattering energy $K_{NN}$ for $\tau$, which
we use in the present study, is defined by the kinetic energy of
the incoming proton in the rest frame of the target proton \cite{Amos}.
Note that the
momentum of the incoming proton to be used in the calculation of $M$
is not the asymptotic momentum $\bs{k}_1$ but the local one
$\bs{k}_1(\eta\bs{R})$.
Similarly, the local momenta ${\bs k}'_1({\bs R})$ and
${\bs k}'_2({\bs R})$ dictate the
final states of the two protons of the transition by $\tau$.
This change in the magnitudes and the directions of the momenta
of the incoming and outgoing protons is referred to as the
refractive effect of distortion, which was shown to play an
essential role in the accurate description of ($p,p'x$) reactions \cite{Ogata}.
Another remark on $M$ of Eq.~(\ref{ME}) is that
$\kappa' \neq \kappa$, hence,
the off-shell matrix elements are necessary.

In foregoing studies with nonrelativistic DWIA
\cite{nrDWIA,AyPuzzle2}, simplification
of $M$ with no theoretical foundation was made as follows.
First, the total local momentum of the two protons was assumed to be
conserved in the transition by $\tau$,
namely, $\bs{\kappa}$ of Eq.~(\ref{kpi})
was replaced by
\beq
\bs{\kappa}^\mr{mc}\equiv
\dfrac{
2{\bs k}_1(\eta{\bs R})
-{\bs k}'_1({\bs R})-{\bs k}'_2({\bs R})
}
{2}.
\eeq
This assumption resulted in the following form of the $T$ matrix:
\bea
T_{m'_{1}m'_{2}m_1M_j}^\mr{mc}
&\equiv&
\int d{\bs R}
\chi_{\mr{f}_{1};{\bs k}'_{1}}^{(-)*}({\bs R})
\chi_{\mr{f}_{2};{\bs k}'_{2}}^{(-)*}({\bs R})
\chi_{\mr{i};{\bs k}_{1}}^{(+)}(\eta {\bs R})
\non \\
&&
\times
\sum_{m,m_2}
(1m{\textstyle\frac{1}{2}}m_2|{\textstyle\frac{3}{2}} M_j)
\phi_{1p\frac{3}{2},m}({\bs R})
\non \\
&&
\times
M_{m'_1m'_2m_1m_2}(\bs{\kappa}',\bs{\kappa}^\mr{mc})
\label{Tmc}.
\eea
Second, the above-mentioned refractive effect of distortion
was neglected, i.e.,
\bea
\bs{\kappa}'
&\to&
\bs{\kappa}'_\mr{as}
\equiv
({\bs k}'_1-{\bs k}'_2)/2,
\non \\
\bs{\kappa}^\mr{mc}
&\to&
\bs{\kappa}^\mr{mc}_\mr{as}
\equiv
(2{\bs k}_1-{\bs k}'_1-{\bs k}'_2)/2.
\eea
The third simplification is the so-called on-shell approximation to $M$,
in which $\kappa'$ is changed to $\kappa$ with the direction
of $\bs{\kappa}'$ fixed. Consequently, the following simplified $T$
matrix was used in previous studies:
\bea
T_{m'_{1}m'_{2}m_1M_j}^\mr{prev}
&\equiv&
\int d{\bs R}
\chi_{\mr{f}_{1};{\bs k}'_{1}}^{(-)*}({\bs R})
\chi_{\mr{f}_{2};{\bs k}'_{2}}^{(-)*}({\bs R})
\chi_{\mr{i};{\bs k}_{1}}^{(+)}(\eta {\bs R})
\non \\
&&
\times
\sum_{m,m_2}
(1m{\textstyle\frac{1}{2}}m_2|{\textstyle\frac{3}{2}} M_j)
\phi_{1p\frac{3}{2},m}({\bs R})
\non \\
&&
\times
M_{m'_1m'_2m_1m_2}
\left(\dfrac{\kappa_\mr{as}^\mr{mc}}{\kappa'_\mr{as}}
\bs{\kappa}'_\mr{as},\bs{\kappa}_\mr{as}^\mr{mc}\right).
\label{Tprev}
\eea
In some cases $NN$ $t$ matrix
instead of $g$ was adopted,
neglecting in-medium effects concerned with the Pauli principle
in the evaluation of $NN$ effective interactions from bare
$NN$ forces. This approximation can be simulated in the present calculation
with the Melbourne $g$ matrix by taking its limit in the free space, i.e.,
$\tau$ with $\rho=0$.

As one sees from Eq.~(\ref{Tmc}),
assumption of the local momentum conservation makes numerical calculation
much simple. We thus first see in Fig.~1 the accuracy of this assumption.
%
%%%%%%%%%%%%%%%%%%%%%%%
%%%  Figure 1
%%%%%%%%%%%%%%%%%%%%%%%
\begin{figure}[htbp]
\begin{center}
\includegraphics[width=0.45\textwidth,clip]{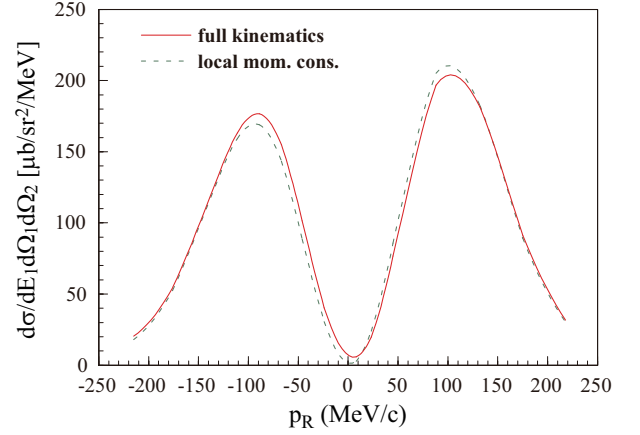}
\caption{
The TDX for the 1p3/2 proton
knockout ($p,2p$) reaction from $^{12}$C at 392 MeV
corresponding to {\it kinematics 1},
as a function of the recoil momentum.
The solid line shows the theoretical result including full kinematics,
while the dotted line shows the
result assuming the local momentum conservation.
In both calculations the spectroscopic factor is set to be 4.
}
\end{center}
\end{figure}
The solid line represents the TDX calculated with $T$ of Eq.~(\ref{Tmat3})
putting $S=4$. The energy and the emission angle of the proton 1 are fixed
at 250 MeV and 32.5$^\circ$, respectively, while the emission angle
of the proton 2 is varied. The azimuthal angle of the detector for the
proton 1 (proton 2) is 0$^\circ$ (180$^\circ$) in the
Madison convention. We refer to this experimental condition as
{\it kinematics 1} following Ref.~\cite{NoroExp}.
The horizontal axis of Fig.~1 is the recoil momentum $p_\mr{R}$
of $^{11}$B that is to be interpreted as the momentum of the
struck target proton in the initial state; the positive (negative)
sign of $p_\mr{R}$ means that
the projection of $\bs{p}_\mr{R}$
onto the $z$-axis (the incident beam direction)
is negative (positive).
Also shown by the dotted line is the same as the solid one but
assuming the local momentum conservation, i.e., using $T^\mr{mc}$
of Eq.~(\ref{Tmc}).
In both calculations
the $p$-$^{12}$C optical potential based on the Dirac phenomenology
\cite{Hama} is
adopted to generate the distorted waves and the proton
single-particle wave function in the 1p3/2 state in $^{12}$C
is calculated with the global potential of Bohr and Mottelson
\cite{BM}.
We will discuss the dependence of the TDX on these potentials below.
The two calculations shown in Fig.~1 agree well with each other, from
which one can conclude that the local momentum conservation is
practically fulfilled in the calculation of the TDX.
Therefore, we henceforth show only the results of TDX assuming
this conservation.

%
%%%%%%%%%%%%%%%%%%%%%%%
%%%  Figure 2
%%%%%%%%%%%%%%%%%%%%%%%
\begin{figure}[htbp]
\begin{center}
\includegraphics[width=0.45\textwidth,clip]{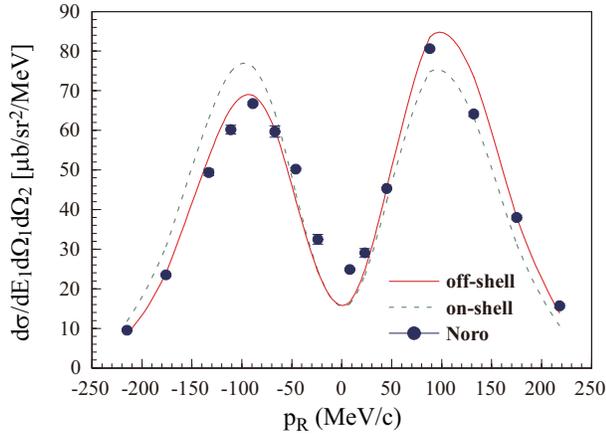}
\caption{
Effects of the off-shell $pp$ scattering on the TDX.
The solid line is the same as the dotted line in Fig.~1 but
with $S=1.71$. The dotted line is the result with the on-shell
approximation to the matrix elements of $\tau$ and $S=1.84$.
The experimental data are taken from Ref.~\cite{NoroExp}.
}
\end{center}
\end{figure}
In Fig.~2 we show the result of the TDX with the $T$ matrix of
Eq.~(\ref{Tmc})
compared with the experimental data \cite{NoroExp};
the theoretical result is averaged over the the emission momentum
$k_1^{\mr L}$ of the proton 1 and the polar and azimuthal angles
of the two detectors, i.e.,
$\theta_1^\mr{L}$, $\phi_1^\mr{L}$, $\theta_2^\mr{L}$ and $\phi_2^\mr{L}$.
The widths for these quantities corresponding to the experiment concerned
are shown in Table I. This smearing procedure is done for all the results
of TDX shown below.
%%%%%%%%%%%%%%%%%%%%%%%
%%%  Table I
%%%%%%%%%%%%%%%%%%%%%%%
\begin{table}[htbp]
\caption{
The widths for $k_1^{\mr L}$, $\theta_1^\mr{L}$,
$\phi_1^\mr{L}$, $\theta_2^\mr{L}$ and $\phi_2^\mr{L}$
used to average the theoretical TDX. See the text for details.
}
\begin{tabular}{cccccccc}
\hline
$\Delta k_1^\mr{L}$\; & $\Delta \theta_1^\mr{L}$\;
& $\Delta \phi_1^\mr{L}$\; & $\Delta \theta_2^\mr{L}$\;
& $\Delta \phi_2^\mr{L}$\;
\\ \hline\hline
$0.022 \times k_1^\mr{L}$\; & 20 [mrad]\;
& 30 [mrad]\; & 50 [mrad]\;
& 45 [mrad]\;
\\ \hline
\end{tabular}
\label{tab1}
\end{table}
The spectroscopic factor $S$ in Eq.~(\ref{TDX}) is determined by the $\chi^2$
fit of the calculated TDX to the experimental data. We obtain
$S=1.71$
that agrees very well with the value determined by
the previous ($e,e'p$) experiment, i.e., $S=1.72 \pm 0.11$ \cite{expS}.
The change of the distorting potential and the proton single-particle
potential adopted is found to have no effects on the shape of the resulting
TDX but bring about change in $S$ of about 10\%.
This uncertainty is,
however, of the same order as of the ($e,e'p$) experimental value of $S$.
The dotted line in Fig.~2 shows the result with the simplified $T$ matrix
of Eq.~(\ref{Tprev}), neglecting the in-medium effects
due to the Pauli principle; the resulting $S$ is 1.84.

One clearly sees that the solid line reproduces the experimental
data \cite{NoroExp}
very well in contrast to the dotted line. It is found by
detailed analysis that the on-shell approximation to $M$ causes
the difference between the two results in Fig.~2; refractive
effects of distortion
and in-medium effects concerned with the Pauli
principle turn out to be negligible in the present analysis of
the TDX.
Thus, it is concluded that the use of the off-shell matrix elements of $\tau$
is necessary to reproduce the experimental data.
In fact, the on-shell approximation to $M$ inevitably results in
the symmetric two peaks of the TDX as shown by the dotted line
in Fig.~2, which can be explained as follows.
If the plane wave approximation is made to the $\chi$
in Eq.~(\ref{Tprev}), one finds
\beq
\dfrac{d^3\sigma}
{dE_1^\mr{L}
d\Omega_1^\mr{L}
d\Omega_2^\mr{L}}
\propto
\left|\tilde{f}_{1p\frac{3}{2},m}(p_\mr{R})\right|^2
\dfrac{d\sigma_{pp}^\mr{on}}{d\Omega},
\label{PWIA}
\eeq
where $d\sigma_{pp}^\mr{on}/d\Omega$ is the on-shell $pp$ elastic cross
section
dictated by the momentum transfer $\bs{q} = \bs{k}_i - \bs{k}_f$.
Since the TDX concerned here is measured under the kinematical condition
so that $\bs{q}$ is fixed at 2.5 fm$^{-1}$,
the TDX calculated with Eq.~(\ref{PWIA})
has the same shape as
$\left|\tilde{f}_{1p\frac{3}{2},m}(p_\mr{R})\right|^2$,
which is an even function of $p_\mr{R}$. Although effects of distortion
slightly affect the shape of the TDX, they never provide it with
the striking asymmetry.
Therefore, it can be said that the asymmetry of the
two peaks of the TDX shown in Fig.~2 is evidence of the off-shell
$pp$ scattering taken place in the ($p,2p$) process.

If one considers in Eq.~(\ref{PWIA}) the off-shell property
of the $pp$ cross section, it depends also on
$\bs{Q} \equiv \bs{k}_i + \bs{k}_f$.
%
%%%%%%%%%%%%%%%%%%%%%%%
%%%  Figure 3
%%%%%%%%%%%%%%%%%%%%%%%
\begin{figure}[htbp]
\begin{center}
\includegraphics[width=0.45\textwidth,clip]{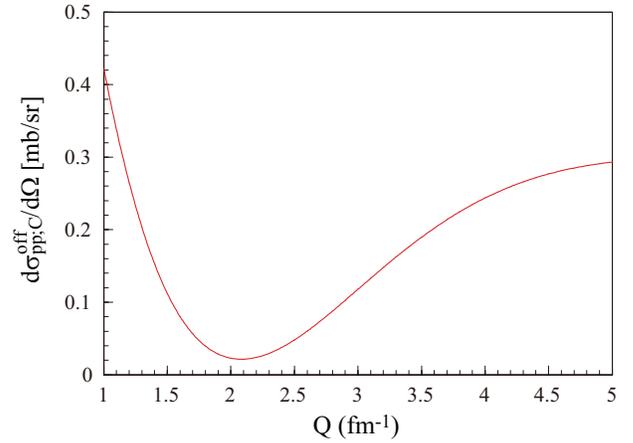}
\caption{
The central part of the off-shell $pp$ cross section as a function of
the exchanged momentum transfer $Q$; the Melbourne $g$ matrix in the
limit of the free space, i.e., at zero Fermi momentum, is adopted.
The momentum transfer $q$ is set to be 2.5 fm$^{-1}$.
}
\end{center}
\end{figure}
In Fig.~3 we show the off-shell cross section of the $pp$ scattering
due to the central part of the Melbourne $g$ matrix,
$d\sigma_{pp;\mr{C}}^\mr{off}/d\Omega$,
as a function of $Q$; $q$ is fixed at 2.5 fm$^{-1}$.
If we adopt the asymptotic kinematics, the value of $Q$ corresponding
to the left (right) peak of the TDX is 3.0 (4.0) fm$^{-1}$.
One sees from Fig.~3 that
$d\sigma_{pp;\mr{C}}^\mr{off}/d\Omega$ for the right peak
is twice as large as that for the left peak.
Thus, the $Q$ dependence of $d\sigma_{pp;\mr{C}}^\mr{off}/d\Omega$
gives the asymmetry of the TDX.
Note that if the on-shell $pp$ scattering is assumed,
$Q$ is uniquely determined by the fixed $q$ as $Q=3.1$ fm$^{-1}$.
Although the contribution of
the central $pp$ interaction to the TDX is only about 20\%,
the change in
$d\sigma_{pp;\mr{C}}^\mr{off}/d\Omega$ due to the off-shell property
shown in Fig.~3 is clearly seen in Fig.~2.

%
%%%%%%%%%%%%%%%%%%%%%%%
%%%  Figure 4
%%%%%%%%%%%%%%%%%%%%%%%
\begin{figure}[htbp]
\begin{center}
\includegraphics[width=0.45\textwidth,clip]{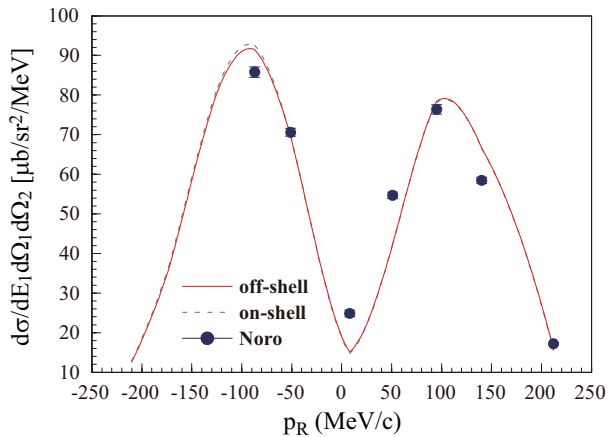}
\caption{
Same as in Fig.~2 but for {\it kinematics 3}.
}
\end{center}
\end{figure}
There are other three sets of TDX data for the proton knockout
($p,2p$) reaction from the 1p3/2 orbit in $^{12}$C, corresponding to
different kinematical conditions \cite{NoroExp}.
We show in Fig.~4 the results of the TDX just in the same way as in Fig.~2.
In the kinematics adopted, which we refer to as {\it kinematics 3},
the energy of the proton 1 is chosen to be 250 MeV and the angle between
the two detectors are almost fixed around 83$^\circ$, i.e.,
79$^\circ$--86$^\circ$.
As discussed in Ref.~\cite{NoroExp}, kinematics 3 was designed to
minimize the off-shell property of the $pp$ scattering in the target nucleus.
This is indeed confirmed by the negligibly small difference
between the two theoretical results shown in Fig.~4;
both results reproduce the experimental data \cite{NoroExp} well.
The value of $S$ corresponding to the solid (dotted) line is
1.64 (1.78).
Other features of the DWIA calculation, i.e., effects of refraction
and in-medium modification to $\tau$ due to the Pauli principle,
as well as
the accuracy of the assumption of the local momentum conservation,
are found to be the same as in the case of kinematics 1.

In summary, the triple differential cross sections (TDX's)
for the 1p3/2 proton knockout ($p,2p$) reaction from $^{12}$C
at 392 MeV are analyzed with nonrelativistic
distorted wave impulse approximation (DWIA)
explicitly taking account of effects of distortion,
the refractive effect in particular, and
those of in-medium modification to the matrix elements $M$ of
the proton-proton ($pp$) effective interaction $\tau$.
We show by numerical calculation
that the conservation of the total momenta
of the colliding two protons is practically fulfilled in the description
of the TDX's of the ($p,2p$) reaction concerned. In-medium modification to
$\tau$ corresponding to the Pauli principle and the refractive
effect of distortion on the kinematics of the incoming and
outgoing protons are both found to have no importance in
the present analysis of the TDX's, which justifies previous
nonrelativistic DWIA calculations in part.
However, explicit treatment of the off-shell
matrix elements of $\tau$ is necessary to reproduce the
experimental data of the TDX with the kinematics in which
the momentum transfer is fixed, i.e., {\it kinematics 1}.
Indeed, the asymmetry of the two peaks of the TDX is reproduced only
when the off-shell matrix elements of $\tau$ are used, hence,
this asymmetry is inferred to be evidence of the off-shell
$pp$ scattering in the ($p,2p$) process.
In the analysis of the TDX with different kinematics in
which the off-shell effects are expected to be small, i.e.,
{\it kinematics 3}, both calculations with and without
the on-shell approximation to $M$ reproduce the experimental
data well. Thus, the successful description of the TDX's with
accurate nonrelativistic DWIA is opening the door to
profound understanding of the reaction mechanism of ($p,2p$)
reactions. Analysis of the spin observables, the analyzing power
$A_y$ in particular, with the present DWBA calculation,
after inclusion of the spin-orbit part of the distorting potential
as in Ref.~\cite{Ogata}, will be very interesting and important.

\begin{acknowledgments}
The authors would like to thank T. Noro and M. Kawai for valuable
discussions.
This work has been supported in part by the Grants-in-Aid
for Scientific Research (Grant No.~17740148)
of the Ministry of Education, Science, Sports, and Culture of Japan.

\end{acknowledgments}

\end{document}